\documentclass{desyproc}
\newcommand\ignore[1]{}
\newcommand\be{\begin{equation}}
\newcommand\ee{\end{equation}}
\newcommand\bea{\begin{eqnarray}}
\newcommand\eea{\end{eqnarray}}

\begin{document}
\title{Elastic and  Diffractive Scattering  after AdS/CFT}


\author{{\slshape Richard Brower~$^1$, Marko Djuri{\'c}~$^2$, Chung-I Tan~$^2$\protect\footnote{\ \ speaker}}\\[1ex]
$^1$Physics Department, Boston University, Boston, MA  02215, USA.\\
$^2$Physics Department, Brown University, Providence, RI 02912, USA.}


\desyproc{DESY-PROC-2009-xx}
\acronym{EDS'09} 

\maketitle

\begin{abstract}
At high energies, elastic hadronic cross sections are believed to be dominated by vacuum
  exchange. In leading order of the leading $1/N_c$ expansion this
  exchange process has been identified as the BFKL {\em Pomeron} or
  its strong AdS dual the closed string
  graviton~\cite{Brower:2006ea}. However difference of particle
  anti-particle cross sections are given by a so-called {\em Odderon}
  carrying C = -1 quantum numbers identified in weak coupling with odd
  numbers of exchanged gluons. Here we show that the dual description
  associates this with the Neveu-Schwartz ($B_{\mu\nu}$) sector of
  closed string theory.  We also discuss the extension of the strong coupling treatment to central diffractive Higgs production at LHC.
\end{abstract}

\section{Introduction}

The subject of near-forward high energy scattering for hadrons has a long history.  The traditional description of high-energy small-angle scattering in
QCD has two components --- a soft Pomeron Regge pole associated with  exchanging  tensor
glueballs, and a hard BFKL Pomeron at weak coupling.
On the basis of gauge/string duality,  a coherent treatment
of the Pomeron was provided \cite{Brower:2006ea}.   
These  results agree with expectations for the
BFKL Pomeron at negative $t$, and with the expected glueball spectrum
at positive $t$, but provide a framework in which they are unified \cite{levintan}. 
Therefore, a firm theoretical foundation  for Pomeron in QCD has been established.
 It is now possible to identify a {\em dual Pomeron} as a well-defined feature of the curved-space string
theory~\cite{Brower:2006ea}.

  We focus here on the recent developments  based on Maldacena's weak/strong duality, (AdS/CFT), relating Yang-Mills
theories to string theories in (deformed) Anti-de Sitter space~\cite{Brower:2006ea,Brower:2007qh,Brower:2009zz,Cornalba:2006xm}. The application of  this duality to Diffractive Scattering and the Pomeron physics represent an important  area where a connection with the string-theory-based techniques can be made. Furthermore, it is now possible to extend this  treatment to central diffractive production of Higgs at LHC.

 In the large 't Hooft coupling limit,  Pomeron can be considered as a  {\em Reggeized Massive Graviton}, propagating in a 5-dimensional curved space, the so-called $AdS_5$, where both the IR (soft) Pomeron and the UV (BFKL) Pomeron are dealt in a unified single step. The connection with the stringy aspects in a five-dimensional description is indeed very direct. In gauge theories with string-theoretical dual descriptions, the Pomeron emerges unambiguously. Indeed, Pomeron is directly related to the graviton and its higher spin partners on the leading (five-dimensional) Regge trajectory.  In AdS/CFT, confinement is associated with a deformed $AdS_5$  geometry having an effective horizon, e.g., that for a blackhole. The solution to this is unknown and represents the major theoretical challenge in model-building. Each model leads to certain unique signature. LHC data can provide guide and direction in this endeavor.

\section{Pomeron and  QCD Parameter-Space}

It is useful to take a step back in examining high energy scattering in QCD. 
From a theoretical stand point it is useful to
consider a 3-parameter space varying the number of colors ($N_c$),
the 't Hooft coupling ($\lambda=g^2 N_c$), and the virtuality of 
an external probe $1/Q$, e.g., that of a virtual photon.

\begin{figure}[bthp]
\quad
\quad
\includegraphics[height=0.2 \textwidth,width=0.4\textwidth]{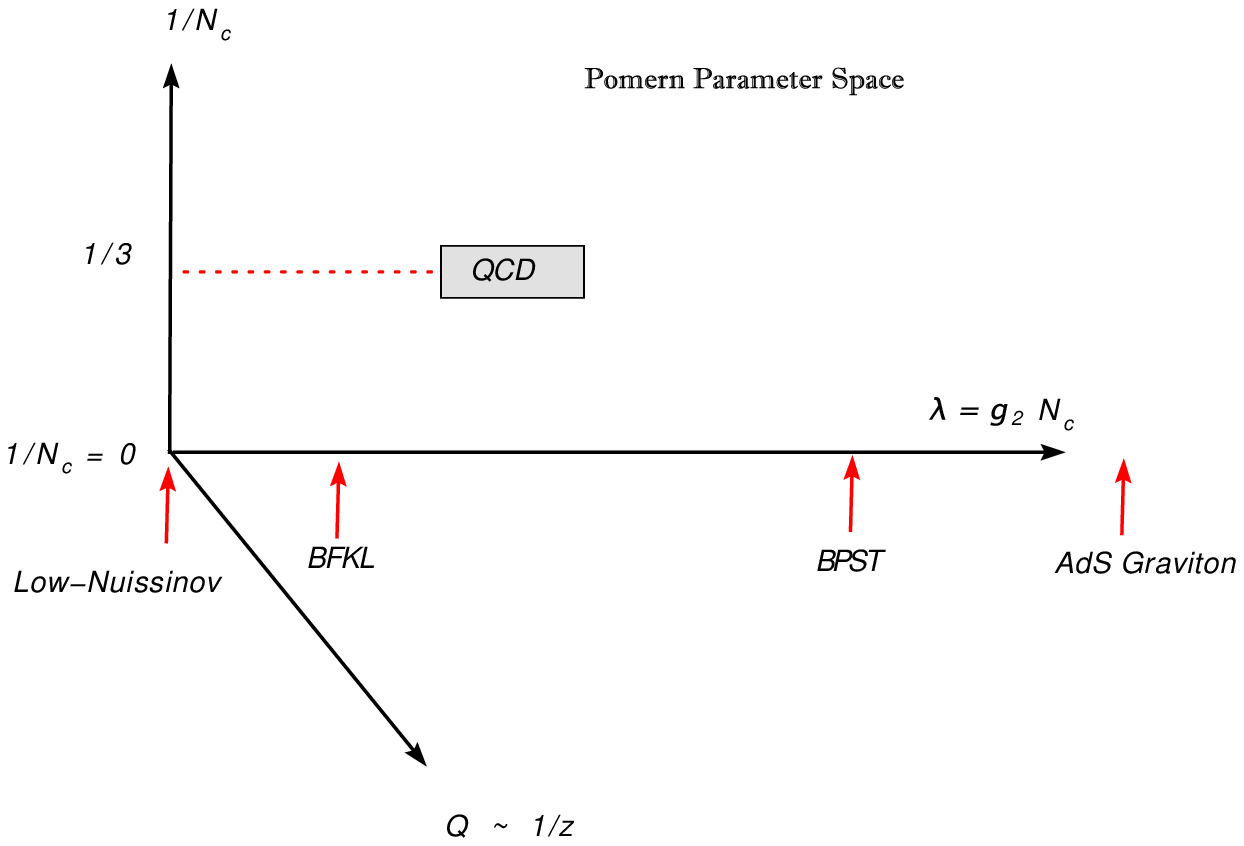}
\quad
\quad
\quad
\quad
\quad
\includegraphics[height=0.15 \textwidth,width=0.35\textwidth]{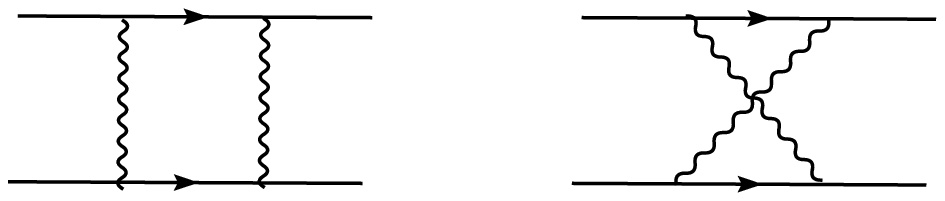}
\quad
\caption{(a) The Pomeron in QCD viewed as a function of
colors ($N_c$), the 't Hooft coupling ($\lambda=g^2N_c$), and the
resolution (virtuality: $Q \sim 1/z$) of the probe. (b) Two-gluon exchange as Low-Nussinov Pomeron.}
\label{fig:4regge}
\end{figure}

Thus diffractive scattering (or Pomeron exchange) in QCD can now be
considered in two steps. First one may consider the leading contribution in the $1/N_c$ expansion holding fixed the 't Hooft coupling $\lambda=g^2N_c$. For example, instead of $N_c=3$,  the leading term for the Regge limit of 2-to-2 scattering in the this limit is the exchange of a network of gluons with the topology of a cylinder in the 't Hooft topological expansion or, in string language, the exchange of a closed string. This gives rise to what we call the ``bare Pomeron'' exchange. Taking into account high order terms in the $1/N_c$ leads to two effects: (1) The cylinder diagrams includes closed quark loops, leading to $q\overline q$ pairs or multi-hadron production via the optical theorem dominated by low mass pions, kaon etc. (2) The multiple exchange of the bare Pomeron which includes the eikonal corrections (or survival probability) and triple-Pomeron and higher order corrections in a Reggeon calculus. We will focus primarily on the ``bare Pomeron'' sector and will discuss only briefly higher order effect due to eikonalization.

 In weak coupling coupling summations where $\lambda<<1$ and $N_c$ large,   the leading singularity  (prior to full unitarization) is at 
$
j_0 = 1 + (\ln 2  /\pi^2)\;  \lambda  \label{eq:BFKL-intecept}
$
where $\lambda = g^2 N_c$ is the 't Hooft coupling. Indeed, in the limit $\lambda\rightarrow 0$, this so-called BFKL Pomeron reduces to the  Low-Nussinov Pomeron, i.e., two-gluon exchange, as depicted in Fig. \ref{fig:4regge}b. When this description is adequate, hadronic cross sections are expected to rise as a small power $s^{j_0-1}$ until unitarity forces compliance with the Froissart bound.  However there is an additional probe of the ``Pomeron'' as a function of virtuality $Q^2$ in off-shell photon scattering.

 It is generally acknowledge that diffractive scattering is intrinsically a non-perturbative phenomena. 
In the limit where the 't Hooft coupling is large,  weak coupling calculations become unreliable. In Ref. \cite{Brower:2006ea}, it has been shown that the leading singularity in strong coupling in the comformal limit approached $j=2$.
In the language of the AdS/CFT, Pomeron is the graviton pole in the 5-dim $AdS$ space where the $AdS$ radius $r$ serves the 5th dimension. (In what follows, this will be referred to as the strong coupling BPST Pomeron, or simply the BPST Pomeron.)

One of the more interesting developments of the BPST Pomeron is the recognition that the virtuality of an external probe, $1/Q$, can be identified with the $AdS$ radius, $z=1/r$.  Conformal invariance, which allows a simultaneous scale transformations in the transverse size and the probe scale, can now be encoded as the isometry of the transverse Euclidean $AdS_3$.

\section{Forward Scattering, Gauge/String Duality, and Confinement}

For conformally invariant gauge theories, the metric of the dual
string theory is a product, $AdS_5 \times W$,  $
ds^2 =\left( \frac{r^2} {R^2}\right) \eta_{\mu\nu} {dx^\mu dx^\nu} +\left(\frac{R^2} {r^2} \right)  {dr^2} + ds^2_W\ ,$
where $0 <r < \infty$.  For the dual to ${\cal N}=4$
supersymmetric Yang-Mills theory  the AdS radius
$R$ is
$R^2 \equiv\sqrt{\lambda} \alpha'= (g_{\rm YM}^2 N)^{1/2} \alpha' \ ,$
and $W$ is a 5-sphere of this same radius.  We will ignore fluctuations over $W$ and also  assume  that $\lambda \gg 1$, so that the spacetime
curvature is small on the string scale, and $g^2_{YM} \ll 1$ so
that we can use string perturbation theory.  (See \cite{Brower:2006ea,Brower:2007qh,Brower:2009zz} for more references.)

\begin{figure}[h]
\quad
\includegraphics[height=0.15 \textwidth,width=0.35\textwidth]{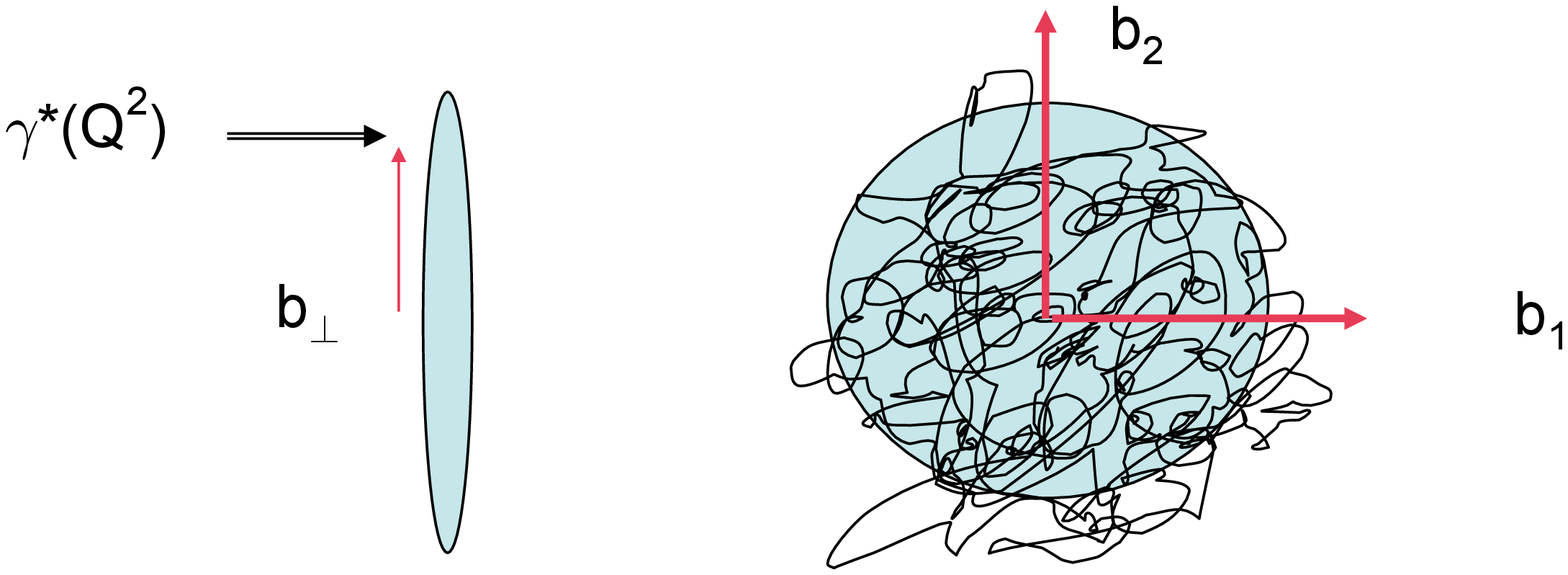}
\qquad
\qquad
\includegraphics[height = 0.15\textwidth,width = 0.25\textwidth]{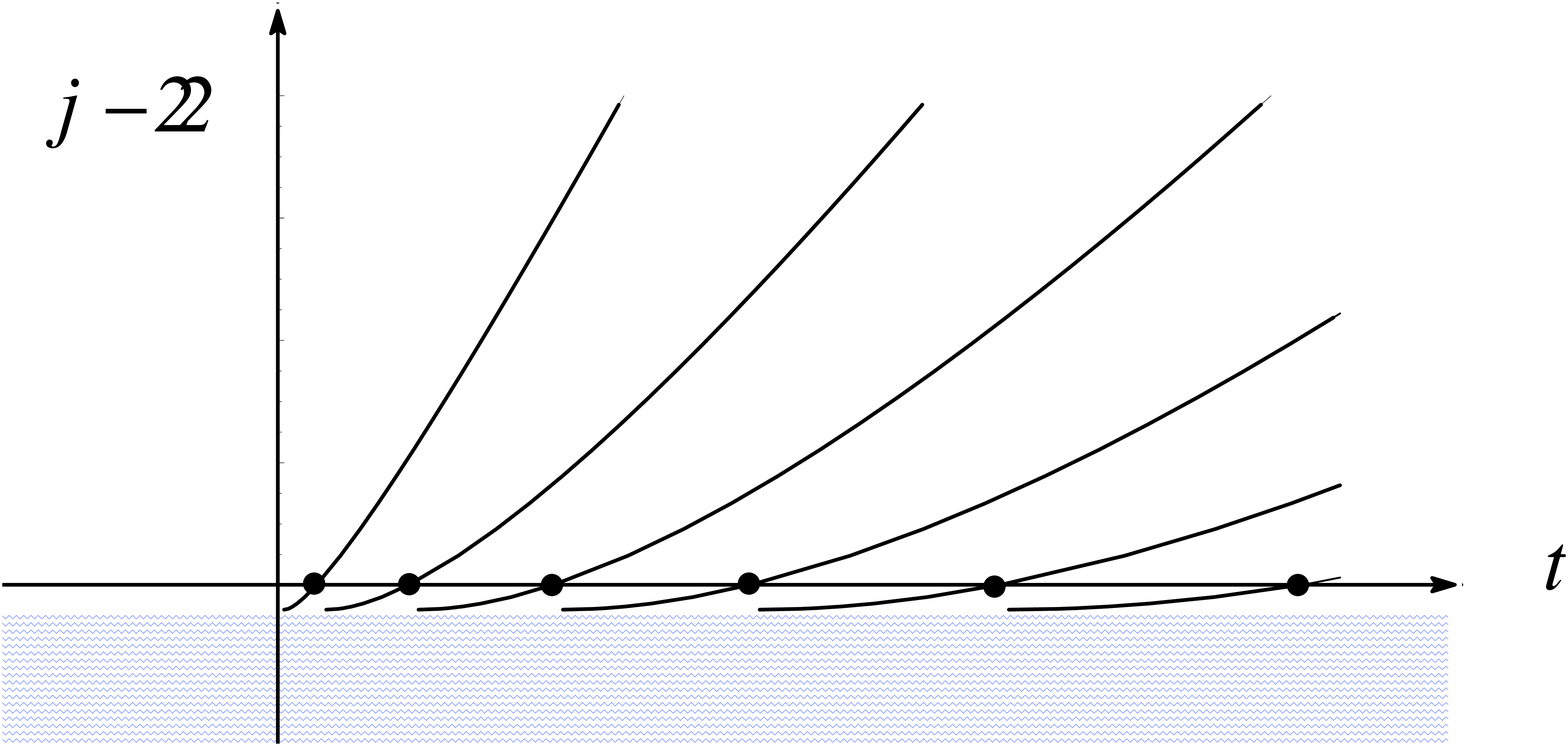}
\qquad
\includegraphics[height = 0.15\textwidth,width = 0.2\textwidth]{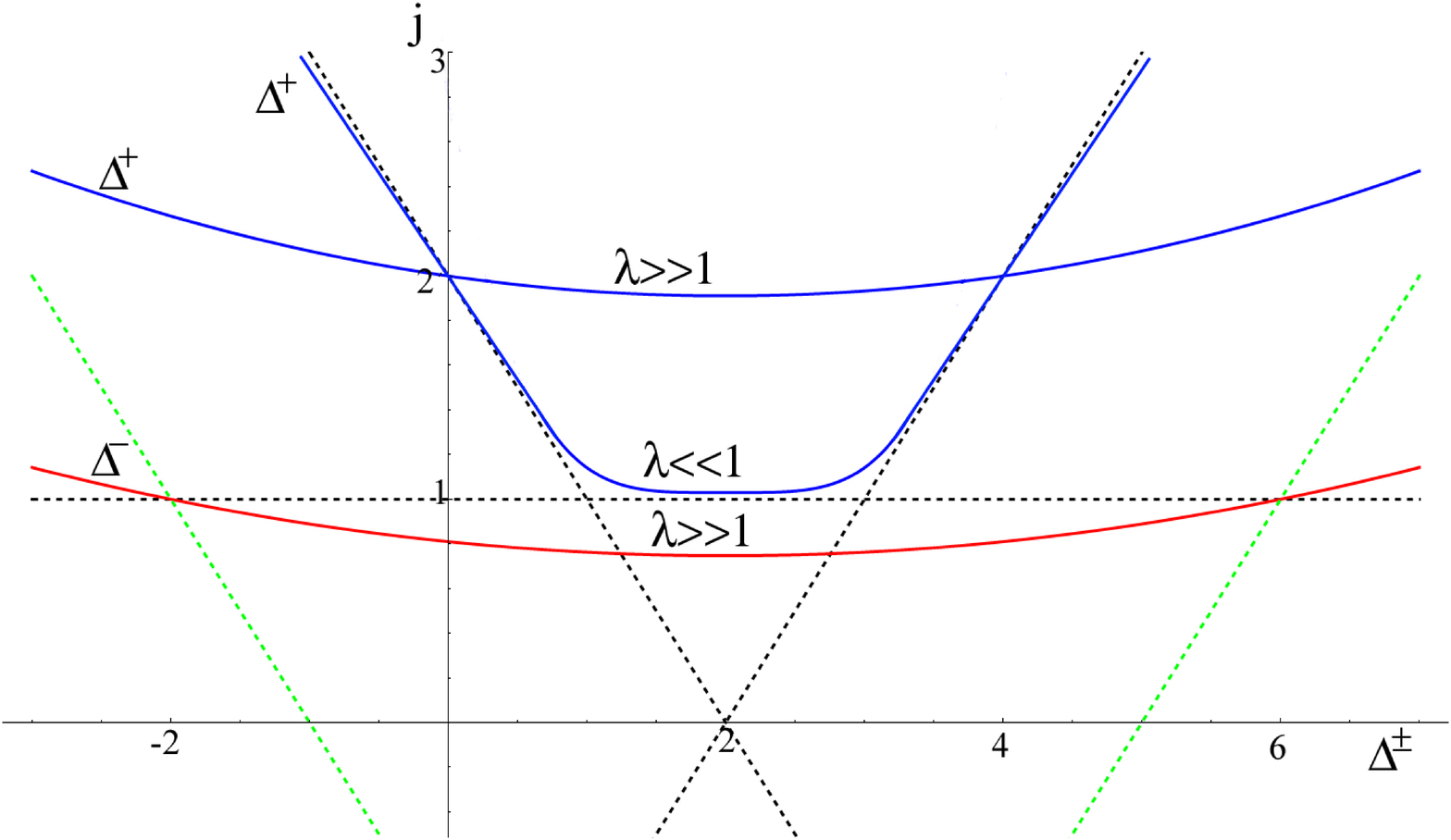}
\quad
 \caption{(a) Intuitive picture for $AdS^5$ kinematics. (b) Schematic representation of $J$-plane singularity structure.
 (c) Schematic form of $\Delta$-$j$ relation for  $\lambda<<1 $ 
  and $\lambda>>1$.}
        \label{fig:comparison}
\end{figure}

The fact that  5-dim description enters in high energy collision can be understood  as follows. In addition to the usual  LC momenta, $p_{\pm}=p^0\pm p^z$ (2d), and transverse impact variables, $\vec b$ (2d), there is  one more ``dimension": a ``resolution" scale specified by a probe, e.g., $1/Q^2$ of virtual photon in DIS, (see  Fig. \ref{fig:comparison}a.) Because of conformal symmetry, these 5 coordinates transform into each others, leaving the system invariant. In the strong coupling limit, conformal symmetry is  realized as the $SL(2,C)$ isometries of Euclidean $AdS_3$ subspace of $AdS_5$, where $r$ can be identified with $Q^2$.

  One important step in  formulating the dual Pomeron involves the demonstration 
\cite{Polchinski:2001tt} that in exclusive hadron scattering, the dual
string theory amplitudes at wide angle, due to the red-shifted local-momenta, $s\rightarrow \tilde s= (R/r)^2 s$ and $t\rightarrow \tilde t= (R/r)^2 t$, give the power laws that are
expected in a gauge theory.   It was also noted that at large $s$
and small $t$ that the classic Regge form of the scattering amplitude should be present in certain kinematic regimes~\cite{Polchinski:2001tt}.     Equally important is the fact  that,  with confinement, transverse fluctuations of the metric tensor $G_{MN}$ in $AdS$ acquire a mass and can be identified with a tensor glueball~\cite{Brower:1999nj,Brower:2000rp}. It was suggested in \cite{Brower:2000rp} that, at finite $\lambda$,   this will  lead to a Pomeron with an intercept below 2. That  is, Pomeron can be considered as a  {\em Reggeized Massive Graviton}.

 For a conformal theory in the large $N_c$ limit, a dual Pomeron can always be identified with the leading eigenvalue of a Lorentz boost generator
$M_{+-}$ of the conformal group  \cite{Brower:2007qh}.  
The problem reduces to finding the spectrum of
a single $J$-plane Schr\"odinger operator.
One finds that, in the strong coupling limit, conformal
symmetry  requires that the leading $C=+1$ Regge
singularity is  a fixed $J$-plane cut,
$
j^{(+)}_0 = 2 - 2 /\sqrt{\lambda}   
$
   For ultraviolet-conformal
theories with confinement deformation,  the spectrum exhibits a set of Regge trajectories at
positive $t$, and a leading $J$-plane cut for negative $t$, the
cross-over point being model-dependent.  (See Fig. \ref{fig:comparison}b.) For theories with
logarithmically-running couplings, one  finds a discrete
spectrum of poles at all $t$, with a set of slowly-varying and closely-spaced
poles at negative $t$.

\section{Conformal Pomeron, Odderon and Analyticity}

At high-energy, analyticity and crossing lead to $C=\pm 1$ vacuum exchanges,  the {\em Pomeron} and  the {\em Odderon}.    The qualitative picture for Pomeron exchange in weak coupling~\cite{Lipatov:1976zz} has been understood for a long time, in leading order expansion 
in $g^2_{YM}$ and all order sum in $g^2_{YM} log(s/s_0)$. In the conformal limit,
both the weak-coupling BFKL Pomeron and Odderons correspond to  $J$-plane branch points, e.g.,  the BFKL Pomeoron is a cut at $
j^{(+)}_0$, above $j=1$.    Two leading Odderons have been identified. (See \cite{Brower:2009zz,Ewerz:2005rg} for more references.)  Both are branch cuts in the $J$-plane.
One has an intercept slightly below 1
\cite{Janik:1998xj}, and the second has an
intercept precisely at 1  \cite{Bartels:1999yt}. 
These are summarized in Table~\ref{tab:intercepts}.  

In the strong coupling limit, as we have already mentioned above, conformal
symmetry
dictates that the leading $C=+1$ Regge
singularity is  a fixed $J$-plane cut at   
$
j^{(+)}_0 = 2 - 2/\sqrt{\lambda}+O(1/\lambda).
$
As $\lambda$ increases, the ``conformal Pomeron" moves  to $j=2$ from below, approaching    the  $AdS$  graviton.  We  have recently shown \cite{Brower:2009zz}   that the strong coupling  {\em conformal odderons} are again
fixed cuts in the $J$-plane, with intercepts specified by the
AdS mass squared, $m^2_{AdS}$, for Kalb-Ramond fields ~\cite{Kalb:1974yc},
$
j_0^{(-)}=1- m^2_{AdS}/2\sqrt{\lambda} + O(1/\lambda)\;. 
$
 Interestingly, two leading {\em dual odderons} can be identified, parallel the weak-coupling situation. One solution has $m^2_{AdS, (1)} = 16$. There is also  a second solution where $m^2_{AdS, (2)}=0$.     We outline below how these features emerge in {\em Gauge/String duality}.

\begin{table}[h]
\begin{center}
\begin{tabular*}{155mm}{@{\extracolsep\fill}||c||l|ll||}
\hline
\hline
 & \multicolumn{1}{c|}{Weak Coupling}   &
 \multicolumn{1}{c}{Strong Coupling}            &  \\
\hline\hline\hline
$C=+1$: Pomeron & $j^{(+)}_0 = 1
+  ( \ln 2) \; \lambda/ {\pi^2}  +O(\lambda^2)$    
   & $j^{(+)}_0 = 2 - 2/\sqrt{\lambda}+O(1/\lambda)$             &     \\
\hline\hline
$C=-1$:  Odderon   &   $j^{(-)}_{0,(1)} \simeq  1- 0.24717\;  \lambda/\pi + O(\lambda^2)$                                      
     & $j^{(-)}_{0,(1)}=1- 8/\sqrt{\lambda} + O(1/\lambda)$                   &     \\
       &     $ j^{(-)}_{0,(2)} = 1 + O(\lambda^3)$              
      & $j^{(-)} _{0,(2)}=1+ O(1/\lambda)$                    & \\
\hline
\hline
\end{tabular*}
\caption{Pomeron and Odderon intercepts at weak and strong coupling. }\label{tab:intercepts}
\end{center}
\end{table}

\subsection{Flat-Space Expectation for $C=\pm 1$ Sectors}

String scattering  in 10-d flat-space at high energy leads to  a crossing-even  and crossing-odd amplitudes, 
$
 {\cal T}^{(\pm)}_{10}(s,  t) \to f^{(\pm)} (\alpha'  t)  (\alpha'  s)^{\alpha_\pm(t)} \;,
$
where   $\alpha_{+} (t) = 2 + \alpha'  t /2$ and $ \alpha_{-}(t) = 1 + \alpha'  t /2$ respectively. That is, at $t=0$, a massless state with integral spin is being exchanged, e.g., for $C=+1$, one is  exchanging a massless spin-2 particle, the ubiquitous graviton.  Of course, the coefficient functions, $ f^{(\pm)} (\alpha'  t) $, are process-dependent.

Massless modes of a closed string theory  can be identified with transverse fluctuations coming from a left-moving and a right-moving level-one oscillators, e.g., states created by applying $  a^\dagger_{1,I}\tilde a_{1,J}^\dagger$ to the  vacuum, i.e., $ a^\dagger_{1,I}\tilde a_{1,J}^\dagger |0;k^+,k_\perp\rangle$,   with $k^2=0$.     Since a 10-dim closed string theory in the low-energy limit  becomes 10-dim gravity; these  modes  can be identified  with fluctuations of the metric $G_{MN}$, the anti-symmetric \underline {Kalb-Ramond} background $B_{MN}$ ~\cite{Kalb:1974yc},  and the dilaton, $\phi$, respectively.    It is  important to note that we will soon focus on $AdS^5$, i.e., one is effectively working at $D=5$. With $D=5$,  the independent components for $G_{MN}$ and $B_{MN}$ are 5 and 3 respectively, precisely that necessary for having  (massive) states with  spin 2 and 1 \cite{Brower:2000rp}. For oriented strings, it can be shown that  the symmetric tensor contributes to $C=+1$ and the anti-symmetric tensor contributes to $C=-1$.

\subsection{Diffusion in AdS  and DGLAP Connection}
\label{sec:diffusion}

 Let us next  introduce diffusion in  AdS.  We will  restrict ourselves  to  the conformal limit.   Regge behavior is intrinsically non-local in the transverse space.  For flat-space scattering in 4-dimension, the transverse space is the 2-dimensional  impact parameter space, $\vec b$. In the Regge limit of $s$ large and  $t<0$, the momentum transfer is transverse. Going to the $\vec b$-space, $ t   \to   \nabla_b^2\;, $
and the flat-space Regge propagator, for both $C=\pm 1 $ sectors,  is nothing but a diffusion kernel,
$  \langle \; \vec b \;|\;(\alpha' s)^{\alpha_{\pm}(0)+\alpha' t\nabla_b^2/2} \;| \; \vec b' \;\rangle $, with  $\alpha_+(0)=2$ and $\alpha_-(0)=1$ respectively. 
In moving to a
ten-dimensional momentum transfer $\tilde t$,   we must keep a term  coming from the momentum transfer in the six transverse directions.  This extra term leads to diffusion in  extra-directions, i.e., for $C=+1$, 
$
\alpha'  \tilde t \to  {\alpha'} \Delta_{P} \equiv
\frac{\alpha' R^2 }{r^2} \nabla_b^2+ {\alpha'} \Delta_{\perp P}.
$
The transverse Laplacian is proportional to $R^{-2}$, so that the added term
is indeed of order ${\alpha'}/{R^2} = 1/\sqrt{\lambda}$. 
To
obtain the $C=+1$ Regge exponents we will have to diagonalize the differential
operator $\Delta_{P}$.  Using a Mellin transform,  $\int_0^{\infty} d\tilde s \; {\tilde s}^{-j-1}$,  the Regge propagator can be expressed as 
$
\tilde s^{ 2+ \alpha' \tilde t /2 }
 = \int \frac{d j}{2\pi i} \; {\tilde s}^j \;    G^{(+)}(j) =  \int \frac{d j}{2\pi i} \; \frac{{\tilde s}^j }  { j-  2 -\alpha' \Delta_P /2 }
$
where  $\Delta_P \simeq  \Delta_j  $,  the tensorial Laplacian.  Using a spectral analysis, it leads to a $J$-plane cut at $j_0^{(+)}$.  

A similar analysis can next be carried out for the $C=-1$ sector. We simply replace the  Regge kernel by 
$
\tilde s^{ 1+ \alpha' \tilde t /2 } =\int \frac{d j}{2\pi i} \; {\tilde s}^j \;    G^{(-)}(j) =   \int \frac{d j}{2\pi i} \; {{\tilde s}^j }   {( j-  1 -\alpha' \Delta_O /2)^{-1} }
$. The operator $\Delta_O(j)$ can be fixed by examining  the EOM at $j=1$ for the associated super-gravity fluctuations responsible for this exchange, i.e., the anti-symmetric Kalb-Ramond fields, $B_{MN}$.  One finds two solutions,
$
G^{(-)}(j) = \frac{1}{[ j-1 -(\alpha'/2R^2) (\square_{Maxwell} - m^2_{AdS,i}) ]}\; ,
$
$i=1,2$, where $\square_{Maxwell}$ stands for the Maxwell operator.  Two allowed values  are
$m^2_{AdS,1}=16 $ and  $m^2_{AdS,2} = 0.$  
 A standard spectral analysis then lead to a branch-cut at $j_0^{(-)}$.

 It is also useful to explore the conformal invariance as the  isometry of  transverse $AdS_3$. Upon taking a two-dimensional Fourier transform with respect to $q_\perp$,  where  $t=-q_\perp^2$, one finds  that $G^{(\pm)}$ can be expressed simply as
\be
G^{(\pm)}(z,x^\perp,z',x'^{\perp}; j)  = \frac{1}{4\pi zz' } \frac{ e^{ (2 -\Delta^{(\pm)}(j))\xi}}{  \sinh \xi} \; ,
\label{eq:adschordal}
\ee
where $\cosh \xi = 1+ v$,   $v= [(x^\perp-x'^{\perp})^2+(z-z')^2]/(2 zz')$ 
the $AdS_3$ chordal distance, and $z=R^2/r$,  and 
$
\Delta^{(\pm)} (j) =  2 + \sqrt {2}\;  \lambda^{1/4}
\sqrt{ (j-j^{(\pm)}_0) }
$
is a  $J$-dependent effective  $AdS_5$ conformal dimension \cite{Brower:2006ea,Brower:2007qh,Brower:2009zz}. The $\Delta-j$ curve  for $\Delta^{(\pm)}$ is shown  in Fig. \ref{fig:comparison}c. 

\section{Unitarity, Absorption, Saturation and the Eikonal Sum}
\label{sec:eikonal}

 For simplicity, we will focus here on the $C=+1$ sector, assuming all crossing odd amplitudes vanish. It has been shown in 
 Refs.~\cite{Brower:2007qh,Cornalba:2006xm} that,
  in the strong coupling limit, a 2-to-2 amplitude, $A(s,t)$, in the near-forward limit  can be expressed in
  terms of a ``generalized'' eikonal representation,
\be
\; A_{2\to2}(s,t) =\int dz dz' P_{13}(z) P_{24}(z') \int d^2b \; e^{-ib^\perp q_\perp}  \widetilde A(s,b^\perp,z,z')\;,
\ee
where 
$
\widetilde A(s,b^\perp,z,z')= 2 i s \left [1- e^{i\chi(s,b^\perp, z,z')} \right] \;,
$
and  $b^\perp = x^\perp - x'^\perp$ due to
translational invariance.    The probability distributions for left-moving, $P_{13}(z)$, and right moving, $P_{14}(z)$ particles are products of initial (in) and final (out)
particle wave functions. The eikonal, $ \chi$,  can be related to  the strong coupling Pomeron kernel
~\cite{Brower:2006ea,Brower:2007qh}, and can be expressed  as  the inverse Mellin transform of $G^{(+)}(j,x^\perp - x'^\perp,z,z')$.

We note the  salient feature of eikonal scattering locally in transverse
$AdS_3$, and the near-forward field-theoretic amplitude is obtained
from a bulk eikonal amplitude after convolution. It is useful to focus
our attention on the properties of the bulk eikonal formula
$\widetilde A(s,b^\perp,z,z')$ itself. For $\chi$
real, it is elastic unitary.  On the other hand, when $\chi$ is
complex, (with ${\rm Im} \chi >0$), one has inelastic production. Absorption and saturation can now be addressed in this context. It is also important to note that, for Froissart bound, confinement is crucial.   Discussion on these and related  issues  can be found in Ref. \cite{Brower:2007qh}. 
 For applications of \cite{Brower:2006ea,Brower:2007qh,Brower:2009zz,Cornalba:2006xm}  for DIS, see \cite{Levin:2009vj}. For a more proper treatment while taking into account of confinement effects, see \cite{BDST}.

\section{Diffractive Production of Higgs at LHC}
A promising production mechanism for Higgs meson at the LHC involves the forward proton-proton scattering $p p \rightarrow p H p$. Because of the exceptional signal to background discrimination, this may even be a discovery channel depending of course on the production cross section. The theoretical estimates generally involve the assumption of perturbative contribution of gluon fusion in the central rapidity region~\cite{Martin:2009ku}. In most estimates the Pomeron is effectively replaced by two-gluon exchange, e.g.,  the Low-Nussinov Pomeron.  In spite of the plausibility of this approach,  there are considerable uncontrolled uncertainties. 

\begin{figure}[h]
\qquad
\includegraphics[height=0.15 \textwidth,width=0.4\textwidth]{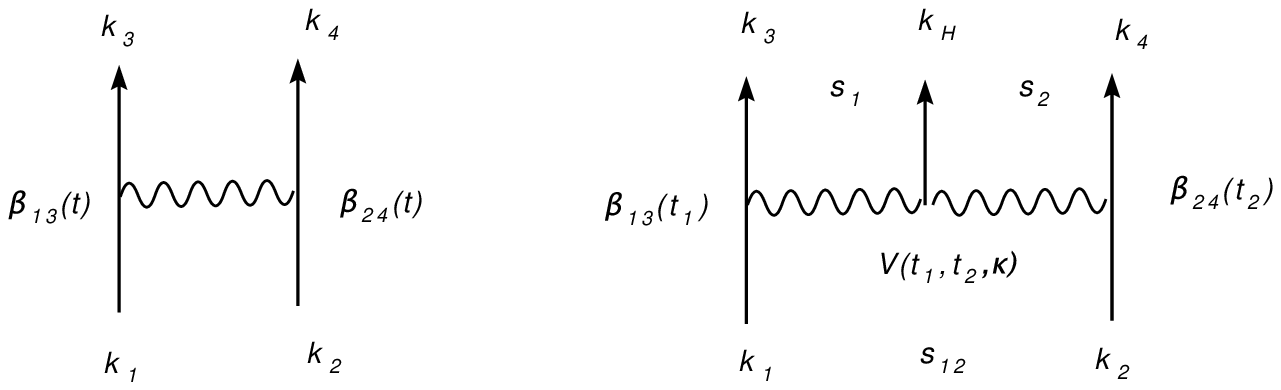}
\qquad
\qquad
\qquad
\includegraphics[angle = 90, height = 0.15\textwidth, width = 0.25\textwidth]{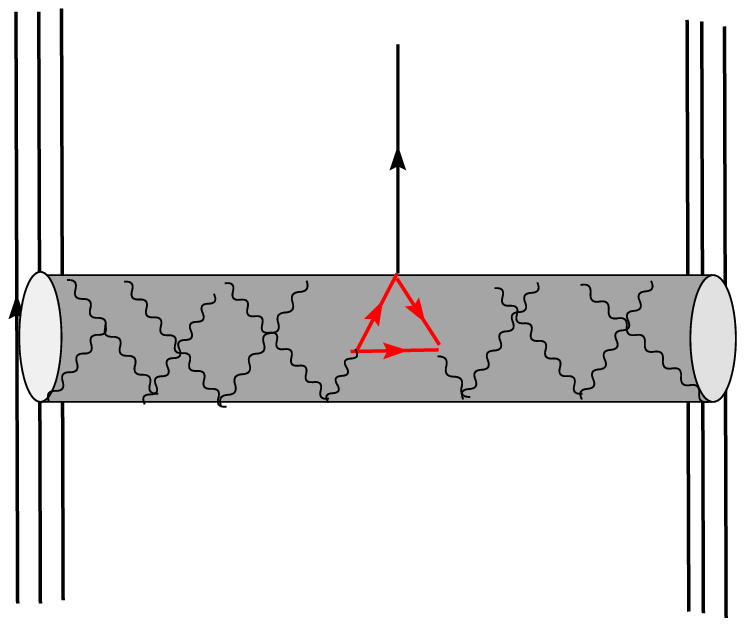}

 \caption{(a) kinematics for single-Regge limit for  2-to-2 amplitudes, (b) Double-Regge kinematics for  2-to-3 amplitudes.  (c) Cylinder Diagram for large $N_c$ Higgs Production.}\label{fig:cylindarHiggs}
\end{figure}
We have begun the analysis in strong coupling based on the AdS/CFT correspondence and conformal  strong coupling BPST Pomeron~\cite{Brower:2006ea}  This amounts to a generalization of our previous $AdS$ for 2-to-2 amplitudes to one for  2-to-3 amplitudes, e.g., from Fig. \ref{fig:cylindarHiggs}a to Fig. \ref{fig:cylindarHiggs}b. A more refined analysis for Higgs production involves a careful treatment for that depicted in Fig. \ref{fig:cylindarHiggs}c. While this also will have its uncertainties, a careful comparison between weak  and strong coupling Pomeron should give better bounds on these uncertainties. Ultimately the strong coupling approach calibrated by comparison with experimental numbers for double diffraction heavy quark production, can provide increasingly reliable estimates for Higgs production.

Focusing only on contributions from Pomeron exchange, a flat-space 2-to-2 amplitude in the Regge limit can be expressed as $
A(s,t)  \simeq \beta_{13}(t)\frac{1 + e^{-i \alpha(t)}}{\sin \pi \alpha(t)} (\alpha' s)^{\alpha(t)} \beta_{24}(t)
$, Fig. \ref{fig:cylindarHiggs}a. 
For a 5-point amplitude, there are five independent Invariants, Fig. \ref{fig:cylindarHiggs}b.  In the kinematic region for diffractive scattering where transverse momenta of all produced particles are limited,  $\kappa\equiv  {s_1s_2/s}$ is fixed, with
$
\kappa \simeq m^2_H + q^2_{\perp},
$
in the frame where incoming particles are longitudinal.   Using a double $J$-plane representation,   In the double-Regge region,  a 2-to-3 amplitude can be expressed  can  be represented using a doulbe-$J$-plane representation, as
$T(s,s_1,s_2, t_1,t_2)
\simeq
 \int_{-i\infty}^{i\infty} \frac{dj_1}{2\pi i}\;   \int_{-i\infty}^{i\infty} \frac{dj_2}{2\pi i}\; \xi(j_1)  \; (\alpha' s_1)^{j_1}  \; \xi(j_2)  \;(\alpha' s_2)^{j_2}    
 \beta_{13}(t_1)G_{j_1}(t_1) {\cal V}(t_1,t_2,\kappa) G_{j_2}(t_2) \beta_{24}(t_2)
$
where $\xi$ is the signature factor and ${\cal V}$ is the  Pomeron-Particle-Pomeoern coupling.  

To move on to $AdS$, we simply need to replace $G_j(t)$ and ${\cal V}(t_1,t_2,\kappa)$ by  corresponding generalizations. The essential new feature  is a new vertex, $
 {\cal V} $, depicted in Fig. \ref{fig:cylindarHiggs}c, appropriate for 
a diffractive central Higgs production~\cite{BDT}. From $m_0\ll  m_H \ll m_t$, the
Higgs vertex is repaced by a source for $F^a_{\mu \nu} F^a_{\mu
  \nu}$ at the boundary of AdS ($z \rightarrow 0$). The standard
AdS/CFT dictionary leads to a bulk to boudary propagator $\Delta(x - x',z)$ for the
interior of AdS to this point so that this vertex can be approximated by a
factorized product.
In a  subsequent analysis we will 
 add corrections
due to (i) conformal symmetry breaking, (ii) the  proton   impact
factor and (iii) eikonal ``survival'' probability to obtain phenomenological results for double Pomeron Higgs production at the LHC. These will be reported in  future publications~\cite{BDT}.

.



\end{document}